\documentclass[prx,twocolumn,floatfix,superscriptaddress]{revtex4}

\usepackage{braket}
\usepackage{siunitx}
\usepackage{hyperref}
\usepackage{multirow}

\usepackage{graphicx}
\usepackage{dcolumn}
\usepackage{bm}
\usepackage{amsmath}
\usepackage{amsfonts}
\usepackage{amssymb}
\usepackage{fancyhdr}
\usepackage{xcolor}

\begin{document}

\title{Probing Environmental Spin Polarization with Superconducting Flux Qubits}
\author{T. Lanting}
\affiliation{D-Wave Systems Inc., 3033 Beta Avenue, Burnaby, British Columbia V5G 4M9,
Canada}
\author{M.~H. Amin\footnote[2]{corresponding author, e-mail: amin@dwavesys.com}}
\affiliation{D-Wave Systems Inc., 3033 Beta Avenue, Burnaby, British Columbia V5G 4M9,
Canada}
\affiliation{Department of Physics, Simon Fraser University, Burnaby BC,
Canada, V5A 1S6}
\author{C. Baron}
\affiliation{D-Wave Systems Inc., 3033 Beta Avenue, Burnaby, British Columbia V5G 4M9,
Canada}
\author{M. Babcock}
\affiliation{D-Wave Systems Inc., 3033 Beta Avenue, Burnaby, British Columbia V5G 4M9,
Canada}
\author{J. Boschee}
\affiliation{Department of ARC, University of British Columbia,
Vancouver, British Columbia V6T 1Z3, Canada}
\author{S. Boixo}
\affiliation{Google Inc., Venice, California 90291, USA}
\author{V.~N. Smelyanskiy}
\affiliation{Google Inc., Venice, California 90291, USA}
\author{M. Foygel}
\affiliation{SGT, Inc., NASA Ames Research Center, Moffett Field, California 94035, USA}
\author{A.~G.~Petukhov}
\affiliation{Google Inc., Santa Barbara, California 93117, USA}
\date{\today}

\begin{abstract}

We present measurements of the dynamics of a polarized magnetic environment
coupled to the flux degree of freedom of rf-SQUID flux qubits. The qubits
are used as both sources of polarizing field and detectors of the environmental polarization. We probe dynamics at timescales from 5\,$\mu$s to 5\,ms and at temperatures between 12.5 and 22 mK. The measured polarization versus temperature provides strong evidence for a phase transition at a temperature of $5.7\pm0.3$ mK. Furthermore, the environmental polarization grows initially as $\sqrt{t}$, consistent with spin diffusion dynamics. However, spin diffusion model deviates from data at long timescales, suggesting that a different phenomenon is responsible for the low-frequency behavior.
A simple $1/f$ model can fit the data at all time scales but it requires empirical low- and high-frequency cutoffs. We argue that these results are consistent with an environment comprised of random clusters of spins, with fast spin diffusion dynamics within the clusters and slow fluctuations of the total moments of the clusters.
\end{abstract}

\maketitle

\section{introduction}

Superconducting qubits are rapidly developing and offer a promising path to a large-scale quantum computing technology \cite{Wendin_2017}. Magnetic flux noise remains a major limitation in these devices and there is an on-going effort to identify and reduce it. Direct experimental measurements reveal a power spectral density that depends on frequency as $1/f^\alpha$ for small $f$, with $\alpha \lesssim 1$ \cite{clarke1976tunnel,koch1983flicker,wellstood1987low,Yoshihara:2006il,Lanting:2009cy,Bialczak:2007dz}. Despite several decades of investigation, the microscopic origin of such a noise is not well understood, although several theories have been proposed \cite{Koch:2007jl,deSousa2007dangling,Faoro:2008ip,Lanting:2014cm,kechedzhi2011fractal,de2014ising,de20191,laforest2015flux}. 
The most likely source of magnetic noise is electron spin defects located in the vicinity of the qubit wiring, specifically in or near the interface between superconducting wiring material and oxide or dielectric layers \cite{Koch:2007jl,Bialczak:2007dz,Sendelbach:2008kr,Lanting:2009cy}. Moreover, the observed \cite{Sendelbach:2008kr} cross-correlation between flux and inductance noise indicates existence of a long-range ferromagnetic order in the spin environment. This observation has led to models based on thermally fluctuating random clusters of spins \cite{kechedzhi2011fractal,de2014ising,de20191}. Spin diffusion is another attractive model that explains some experimental observations \cite{Faoro:2008ip,Lanting:2014cm}. However, spin diffusion does not predict the observed $1/f$ noise spectrum, over a wide range of frequencies. Especially, at frequencies exceeding 1~kHz, the dependence is predicted to be $f^{-3/2}$~\cite{Lanting:2014cm}. These frequencies are typically out of reach for direct flux noise measurements, which become challenging for timescales shorter than 1~ms. 

\begin{figure}[tp]
\includegraphics[width=0.75\linewidth]{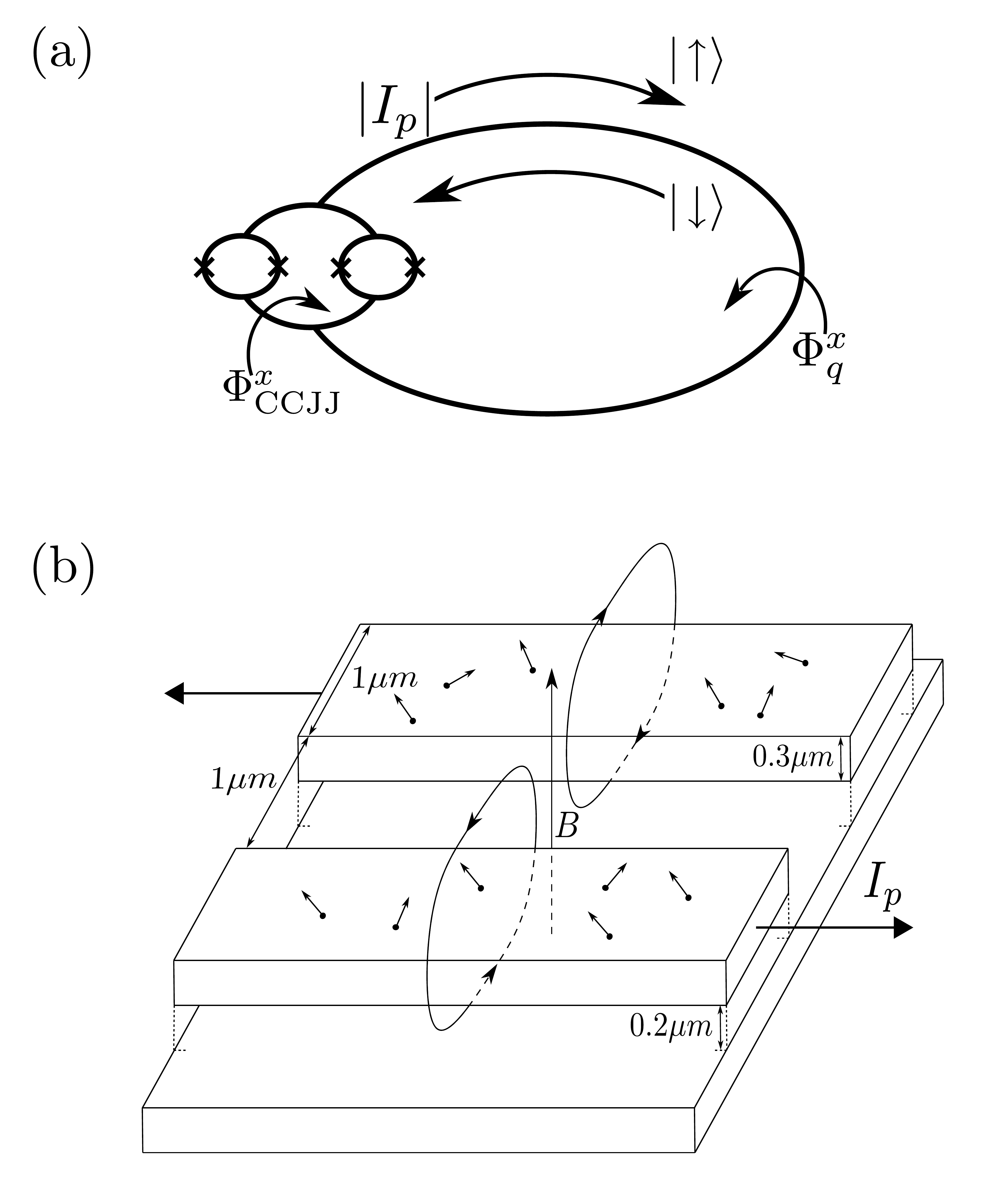}
\caption{(a) Schematic of the rf-SQUID flux qubit used in this study. External control biases $\Phi_{\rm CCJJ}^x$ and $\Phi_q^x$ allow us to adjust parameters $\Delta$, $I_p$, and $\epsilon$ as described in detail in ~\cite{Harris:2010ch}. (b) Cross section of the qubit wiring. The dielectric layer between wiring layers is not shown. Persistent current flowing in the main qubit loop produces a magnetic field that causes a Zeeman splitting of nearby electronic spin defects. At low temperatures, the environmental spins begin aligning with this field.}
\label{FIG:SpinPolarizationQubit}
\end{figure}

The interaction between the magnetic field produced by persistent current flowing in the body of a flux qubit and the surrounding spin environment offers a powerful new way of probing the dynamical behaviour of this environment at shorter timescales. This persistent current produces a magnetic field that causes a fraction of spins to align. This results in a polarization of the environment, which produces a change in magnetic flux bias that shifts the qubit degeneracy point. By varying the time during which persistent current is either present or absent, we can probe the dynamics of the environmental spin polarization and depolarization. 

\begin{figure*}[tp]
\centering
\includegraphics[width=0.85
\textwidth]{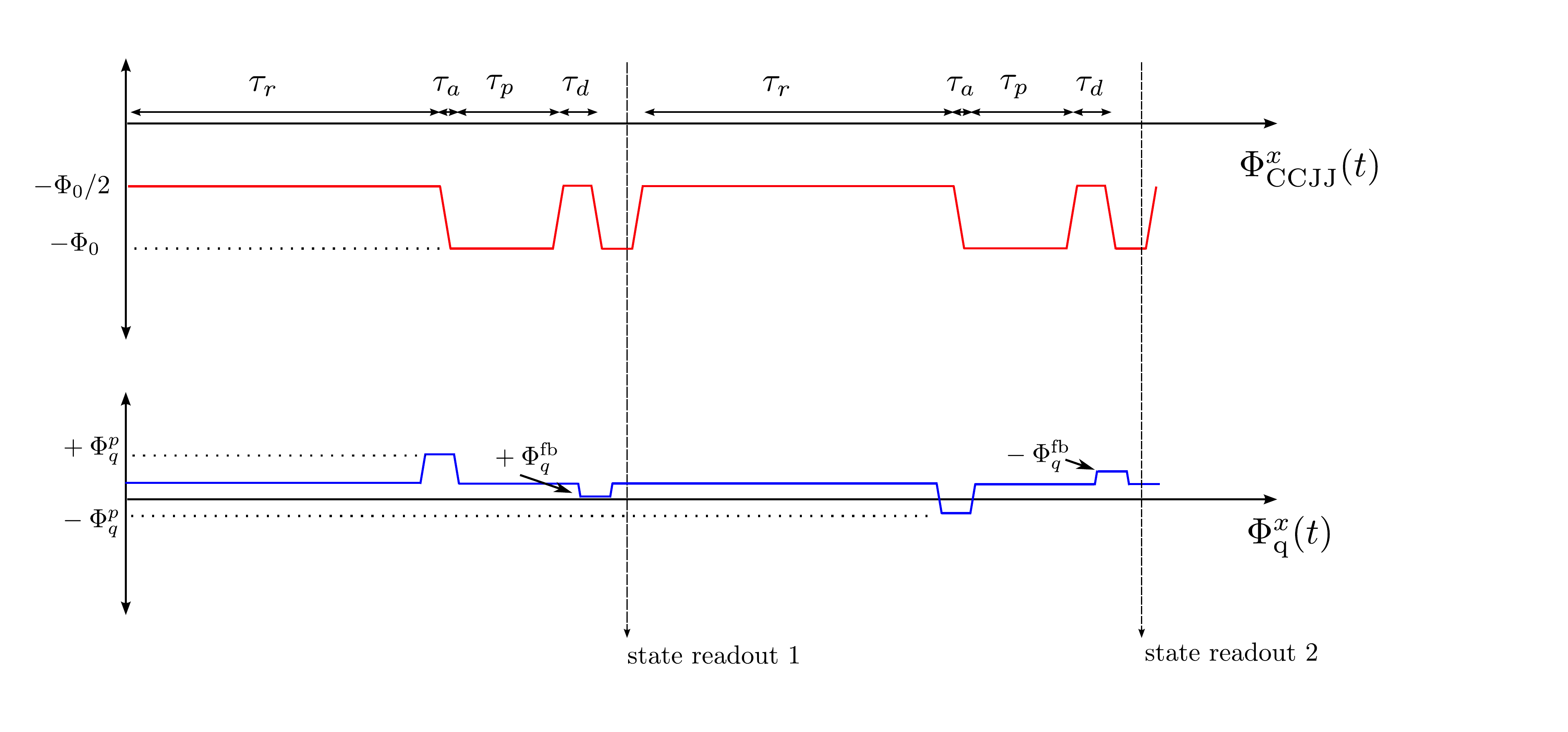}
\caption{Experimental protocol for measuring the polarization of the spin environment. We plot the time-dependent external qubit biases, $\Phi_{\rm CCJJ}^x(t)$ and $\Phi _{q}^{x}(t)$ in the top and bottom panels, respectively. $\Phi_q^x$ is measured with respect to qubit degeneracy. Two opposite
initializations are interleaved to minimize the contribution of low-frequency noise to the measurement. A large ensemble of measurements is performed to determine the feedback signal $\Phi_q^{\rm fb}$ necessary to zero population differences between the two initializations. $\Phi_q^{\rm fb}$ is thus a direct measurement of the magnitude of the spin environment bias $\bar \xi(t)$ on the qubit body.}
\label{FIG:Waveform}
\end{figure*}

Here, we present measurements of environmental spin polarization and depolarization for timescales from 5~$\mu$s to 5~ms and for temperatures ranging from $T=12.5$~mK to 22~mK. We first describe the detailed protocol used for the experiment and then present fits of the dynamics to candidate models. The fits suggest that a random spin diffusion model works well for describing the short timescale growth that shows a $\sqrt{t}$ dependence, but fails to describe the long-time behaviour. The amplitude of the polarization as a function of environment temperature fits well to a Curie-Weiss model with a phase transition at $5.7 \pm 0.3$ mK.

\section{Experiment}

The qubit design used for these experiments is a compound-compound Josephson junctions (CCJJ) rf-SQUID flux qubit~\cite{Harris:2010ch}. Two external control bias lines, $\Phi_q^x$ and $\Phi_{\mathrm{CCJJ}}^x$, shown in Fig.~\ref{FIG:SpinPolarizationQubit}(a) allow control of the qubit dynamics and energy landscape. The Hamiltonian of this qubit coupled to a magnetic environment can be written as:
\begin{equation}
\mathcal{\hat{H}}=-\frac{\Delta}{2}\hat{{\sigma}}_{x}-\frac{1}{2}\left(\epsilon + \xi
\right) \hat{{\sigma}}_{z},
\label{Eq:Ham-qubit}
\end{equation}%
where $\hat{\sigma}_{z,x}$ are the Pauli matrices, $\Delta$ is the tunneling energy,  $\epsilon=2I_{p}\Phi_{q}^{x}$ is the external energy bias, and $I_p$ is the persistent current. The external flux $\Phi_q^x$ is measured relative to the degeneracy point, where the two localized states $\ket{\uparrow}$ and $\ket{\downarrow}$ are equally populated.  Both $I_p$ and the tunneling energy $\Delta$ can be tuned with the external bias $\Phi _{\mathrm{CCJJ}}^{x}$. 
The potential energy of the rf-SQUID can be made monostable, with zero persistent current flowing in the main body, as well as bistable, with nonzero persistent current. The persistent current applies a polarizing magnetic field to the  spins near the surface of the qubit wiring as depicted in  Fig.~\ref{FIG:SpinPolarizationQubit}(b). We treat the environment as an ensemble of classical spins causing a fluctuating energy bias $\xi$. Fast random fluctuations of $\xi$ capture the effect of flux noise with a slow drift of the expectation value $\bar \xi(t)$ (the order parameter) representing polarization or depolarization of the environment.

To measure the environmental polarization, we use the protocol shown in Fig.~\ref{FIG:Waveform}. We begin by adjusting $\Phi_q^x = 0$ ($\epsilon = 0$) and following a two-part protocol. In the first part, the external bias $\Phi _{\mathrm{CCJJ}}^{x}=-0.5\Phi _{0}$ is applied during a recovery time $\tau _{r}$, making the qubit monostable. At this applied bias, the tunneling energy $\Delta\gg k_{\mathrm{B}}T$ while the
persistent current $I_{p}$ is negligibly small, so the qubit and
spin environment are effectively decoupled and each can relax independently. We then initialize the
qubit in one of its two localized states, characterized by a
persistent current $I_{p}$ ($\sim2\ \mu \mathrm{A}$). This is done by applying a
preparation bias $\Phi _{q}^{p}$ while annealing the qubit by changing $\Phi _{\mathrm{CCJJ}}^{x}$ to $-1.0 \, \Phi _{0}$
within anneal time $\tau _{a}$. The qubit potential barrier then stays high for a polarization
time $\tau _{p}$, with effectively no
tunneling between the two bistable states. In this polarization phase, the persistent
current in the qubit body generates a magnetic field that partially polarizes the spin
environment surrounding the qubit wiring.

After the polarization phase, we measure the environmental polarization using the qubit itself. To do this we first need to get the qubit out of its locked state by lowering its energy barrier. We adjust $\Phi _{\mathrm{CCJJ}}^{x}=-0.5\Phi _{0}$ for a short time $\tau _{d}$, enough for the qubit to lose its memory. Once the barrier is raised again, the qubit will be localized into one of its bistable states depending on the direction of the environmental polarization. The state of the qubit at the end of this anneal is measured and recorded. We apply a feedback flux bias $-\Phi _{q}^{\mathrm{fb}}$ to the qubit body before raising the potential barrier and tune it so that the qubit is pushed back to its degeneracy. The magnitude of $\Phi _{q}^{\mathrm{fb}}$ needed to make the qubit have equal population of both states is a direct measure of the polarization flux. The second part of the protocol in Fig.~\ref{FIG:Waveform} is a repetition of the first part 
except the sign of $\Phi _{q}^{p}$ is
reversed. This initializes the qubit in the opposite persistent current state, flipping the direction of the magnetic field polarizing the spin environment, which in turn changes the sign of $\Phi _{q}^{\mathrm{fb}}$. The full protocol is repeated several times and the difference between the results from the two subsequent readouts is
recorded. From this feedback signal we can thus directly determine $|\bar \xi (t_p)|=2|I_p \Phi_q^{\mathrm{fb}}|$.
Note that with the energy barrier low (monostable), the qubit persistent current vanishes, and the qubit decouples from the spin environment. Thus, during time $\tau _{d}$ the spin
polarization starts to relax. To detect the polarization of the spin environment, we typically adjust the protocol such that $\tau _{d}\ll \tau_{p}$. For the depolarization measurement, on the other hand, we allow large values $\tau_d$, while keeping $\tau_p$ fixed.

We applied the protocol shown in Fig.~\ref{FIG:Waveform} to flux qubits in a calibrated quantum annealing processor with 2013 working qubits. Typical polarization and depolarization measurements of $\Phi_{q}^{\mathrm{fb}}$ at $T = 12.5$ mK are shown in Figs.~\ref{FIG:SpinPolarizationTime}(a) and (b). The symbols show the mean signal across all devices in the processor. 
For the polarization growth experiment, plotted in Figs.~\ref{FIG:SpinPolarizationTime}(a), $\tau_p$ is varied while $\tau_d$ is fixed at a small value (1 $\mu$s). For depolarization measurement, we fix $\tau_p$ and measure $\Phi_q^{\rm fb}$ as a function of $\tau_d$, as depicted in Fig.~\ref{FIG:SpinPolarizationTime}(b).

\section{Candidate Models}

\begin{figure}[tbp]
\includegraphics[width=0.95\linewidth]{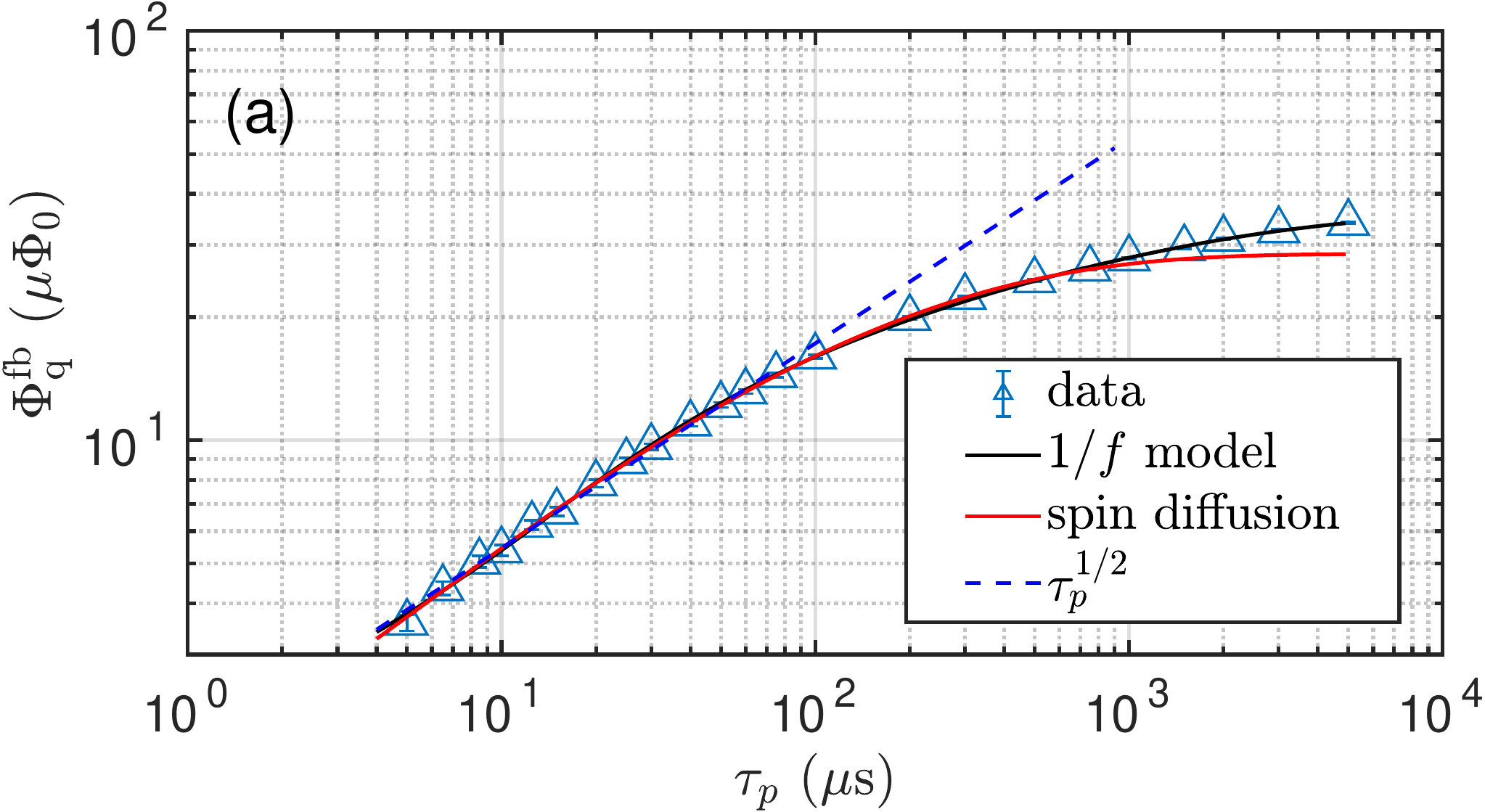}
\includegraphics[width=0.95\linewidth]{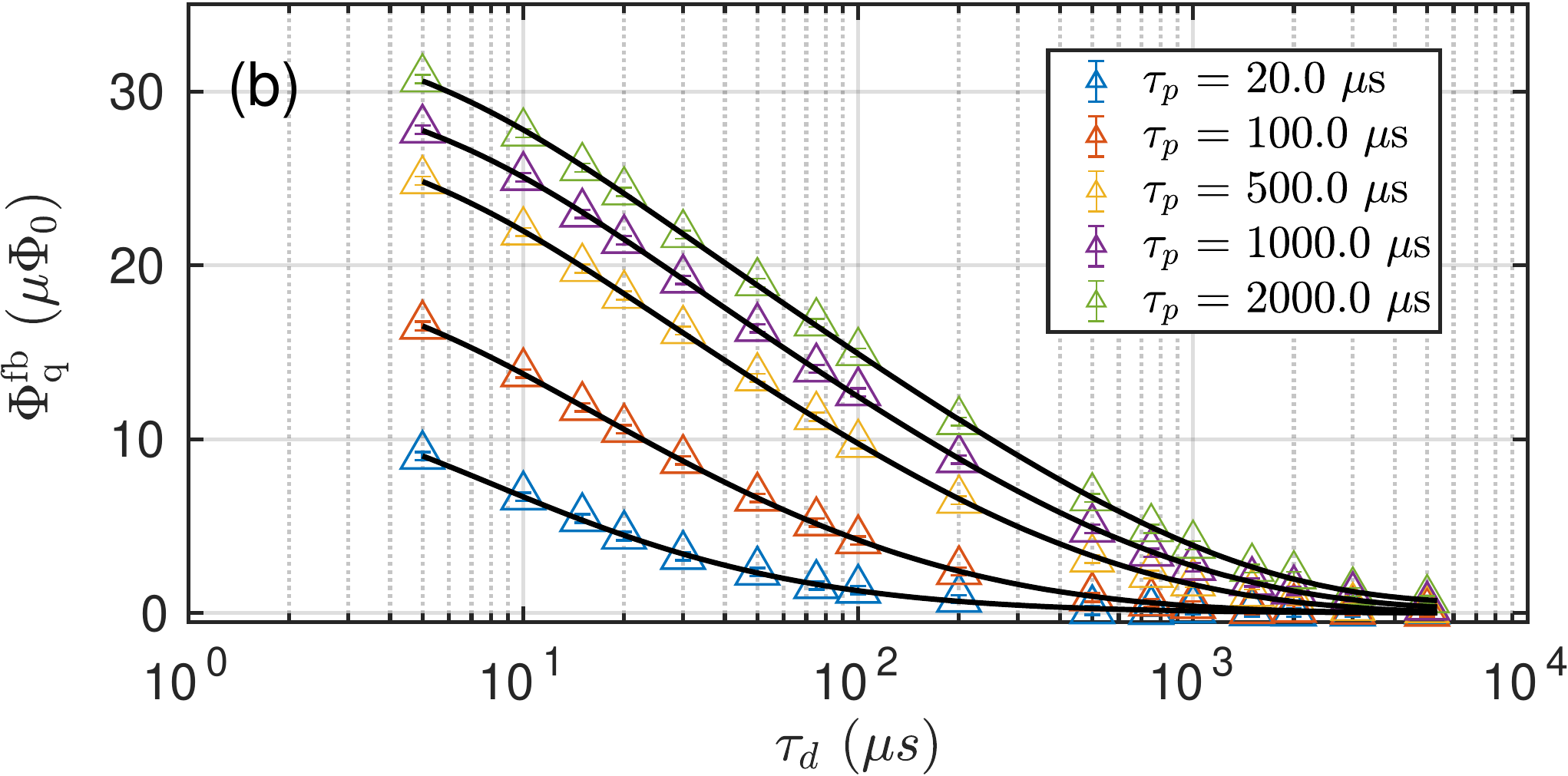}
\caption{Measurements of $\Phi _{q}^{\rm fb}$ versus time for (a) polarization and (b) depolarization experiments at $T = 12.5$ mK. In panel (a) $\tau_d$ is fixed at 1\,$\mu$s.
In both panels, the solid black curves show best fits to Eq.~\eqref{Eq:DeltaPhi1} assuming dynamics governed by the empirical model given by Eq.~\eqref{xi-1/f}. The dashed line in (a) represents $\sqrt{\tau_p}$-dependence and the solid red curve is obtained from spin diffusion model (Eq.~\eqref{Eq:F(t)}) as described in the text.
}
\label{FIG:SpinPolarizationTime}
\end{figure}

To understand the data shown in Fig.~\ref{FIG:SpinPolarizationTime}, we need models that relate measurements of $\Phi_q^{\rm fb}$ to the dynamics of the ensemble of environmental spins that produces the term $\xi$ in Hamiltonian (\ref{Eq:Ham-qubit}).  As we show in Appendix \ref{AppA} (see also~\cite{Kogan:1996}), linear response theory requires a close relation between time-dependent expectation $\bar \xi(t)$, i.e., linear response, and the noise spectral density:
\begin{align} \label{SPhi}
S_\Phi(\omega) =  {T \over \omega I_p^2} \int_{0}^{\infty }dt\sin(\omega t) \left\vert{d\bar \xi (t) \over dt} \right\vert,
\end{align}%
where $\omega=2\pi f$ is the angular frequency.

In the absence of coupling between the qubit and the environment, the environmental spins are in a disordered paramagnetic 
state with zero net magnetization, leading to a zero ensemble average: $\bar \xi(t) = 0$.
The classical states of the qubit are eigenstates of $\hat\sigma_z$.  In these states, the qubit applies a polarizing field to the environment. This produces a nonzero average $\bar \xi(t)$ that is expected to monotonically increase with time until it saturates at its equilibrium value $\epsilon_p \equiv \bar \xi(\infty)$. As soon as the qubit-environment coupling is turned off, the environment starts relaxing back toward $\bar \xi(t) = 0$. 

To model the experimental data, we consider the general case where the environment polarizes within time $\tau_p$ and then depolarizes within time $\tau_d$. In the appendices, we provide detailed derivations for the time dependence of $\bar \xi(t)$ based on several underlying models. For all models, we can write the time dependence as
\begin{align} \label{xi-depol}
\bar\xi(\tau_p,\tau_d) = \epsilon_p [F(\tau_d) - F(\tau_d+\tau_p)].
\end{align}%
For example, to model a polarization experiment, we set $\tau_d\sim 0$ and vary $\tau_p$, and for depolarization we fix $\tau_p$ and study the $\tau_d$ dependence of $\bar\xi$. The envelope function $F(t)$ captures the time dynamics and depends on the specific model of the spin environment. $F(t)$ has the properties: $F(0)=1$ and $F(\infty)=0$. This function fully describes the relaxation behavior and is closely related to the noise spectral density (see Appendix \ref{Sec:Pgamma}) \footnote{A simple exponential function, $F(t) = e^{-\gamma t}$, gives the expected exponential growth, $\bar \xi(\tau_p) = \epsilon_p(1-e^{-\gamma \tau_p})$, and exponential decay $\bar \xi(\tau_d) = \bar \xi(0)e^{-\gamma \tau_d}$, but Eq.~\eqref{xi-depol} holds for any other functional form.}. In the appendices, we consider three different models for the spin environment: homogeneous and inhomogeneous spin diffusion models and a model based on $1/f$ noise spectrum. 

For the spin diffusion model, dynamics is governed by random walk in the space of spin configurations, keeping the total magnetization constant. We consider  homogeneous and inhomogeneous environments. In the homogeneous case, both the distribution of spins and their coupling are assumed to be uniform. In the inhomogeneous case, on the other hand, we assume spins form clusters of random sizes with strong spin-spin interaction within each cluster. Both models are shown to have asymptotic behavior for short and long times given by (see Appendix \ref{app:spindiffusion})
\begin{equation} \label{f-sdmodel}
F(t) = \left\{ \begin{array}{cl} 
1 - C \sqrt{\Omega t} & \text{  for } t \ll \Omega^{-1} \\
C'  e^{-\kappa (\Omega t)^\nu} & \text{  for } t \gg \Omega^{-1}
\end{array}
\right. ,
\end{equation}%
where $C$ and $C'$ are model dependent coefficients and $\kappa = 4 \, (3)$ and $\nu = 1 \, (1/3)$ for the homogeneous (inhomogeneous) case. The parameter $\Omega^{-1}$ is the timescale over which the magnetization can diffuse freely before encountering the geometric boundaries of either the qubit wiring (the homogeneous case) or the clusters (the inhomogeneous case). Notice that in both cases, the short time environmental polarization has the $\sqrt{t}$ dependence expected for random walk. This dependence is associated with asymptotic $f^{-3/2}$ behavior of the noise spectral density through Eq.~\eqref{SPhi}. At long times, on the other hand, the decay is exponential in the homogeneous case and stretched-exponential in the inhomogeneous case. As we shall see in the next section, while the short-time behavior agrees very well with $\sqrt{t}$-dependence, both the exponential or the stretched-exponential decays predict the polarization to saturate at long timescales faster than what is observed experimentally.

We also consider an empirical model for the spin environment assuming a noise power spectral density $S_\Phi(f) = A/f^{\alpha}$, consistent with direct low-frequency observations~\cite{Lanting:2014cm}. We assert a short and long time cutoffs, $\tau_{\rm{min}}$ and $\tau_{\rm{max}}$ such that $S(f < 1/2\pi\tau_{\rm max}) = 0$ and $S(f > 1/2\pi\tau_{\rm min}) = 0$. In Appendix~\ref{app:logarithmic} we show that for this model,
\begin{align}
F(t) = {\cal N} t^{\alpha-1} [\Gamma(1{-}\alpha , \, t/\tau_{\text{max}}) - \Gamma(1{-}\alpha , \, t/\tau_{\text{min}})],
\label{xi-1/f}
\end{align}%
where $\Gamma(s,t)$ is the incomplete gamma function and
\begin{equation}
{\cal N}^{-1} = \left\{ 
\begin{array}{cc}
 \log( \tau _{\text{max}}/\tau _{\text{min}})  & \qquad \alpha = 1 \ \\
(\alpha{-}1)^{-1} [\tau_{\text{max}}^{\alpha-1} - \tau_{\text{min}}^{\alpha-1}] & \qquad  \alpha \ne 1
\end{array}%
\right. 
\end{equation}%
is a normalization factor. Note that Eq.~\eqref{xi-1/f} has three fitting parameters ($\alpha,\tau_{\rm min},\tau_{\rm max}$) in contrast to one ($\Omega$) in the spin diffusion model. In the next section, we explore how these theoretical models fit the experimental data.


\section{Data Analysis}

To fit our experimental data, we express Eq.~(\ref{xi-depol}) directly in terms of flux
\begin{equation}
\Phi _{q}^{\rm fb}(\tau_p)= \Phi_p [F( \tau_d)-F( \tau_d+\tau_p) ] ,  \label{Eq:DeltaPhi1}
\end{equation}%
where $\Phi_q^{\rm fb}$ is the flux bias applied to the body of the qubit by the polarized spin environment, and $\Phi_p=\epsilon_{p}/2 I_p$ is the equilibrium polarization flux. The dashed blue line in  Fig.~\ref{FIG:SpinPolarizationTime}~(a) represents the $\sqrt{\tau_p}$ growth predicted by the short-time limit of the spin diffusion model in Eq.~\eqref{f-sdmodel}. It is clear that the spin diffusion model fits the experimental data for short time scales $\tau_p < $ 1 ms. The red solid line in Fig.~\ref{FIG:SpinPolarizationTime}~(a) is obtained by fitting the inhomogeneous spin diffusion model (Eq.~\eqref{Eq:F(t)}) to experimental data up to $\tau_p = 1$\,ms. Since the cutoff point at 1\, ms is not well defined, the fitting parameters cannot be accurately  determined, but just roughly estimated ($\Phi_p \sim 34 \, \mu\Phi_0$ and $\Omega \sim 100$\,Hz). 
 
\begin{figure}[t]
\includegraphics[width=0.90\linewidth]{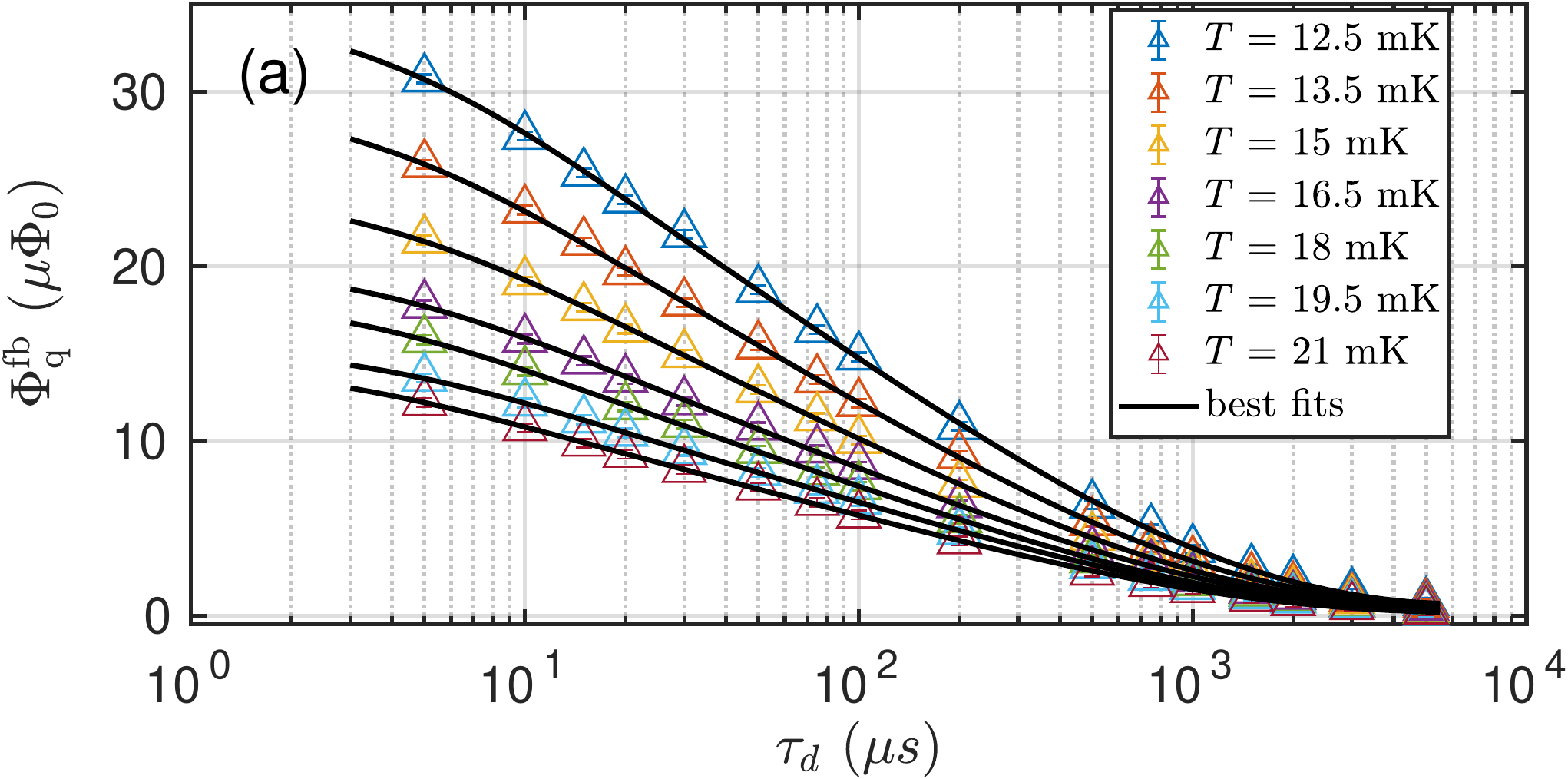}
\includegraphics[width=0.90\linewidth]{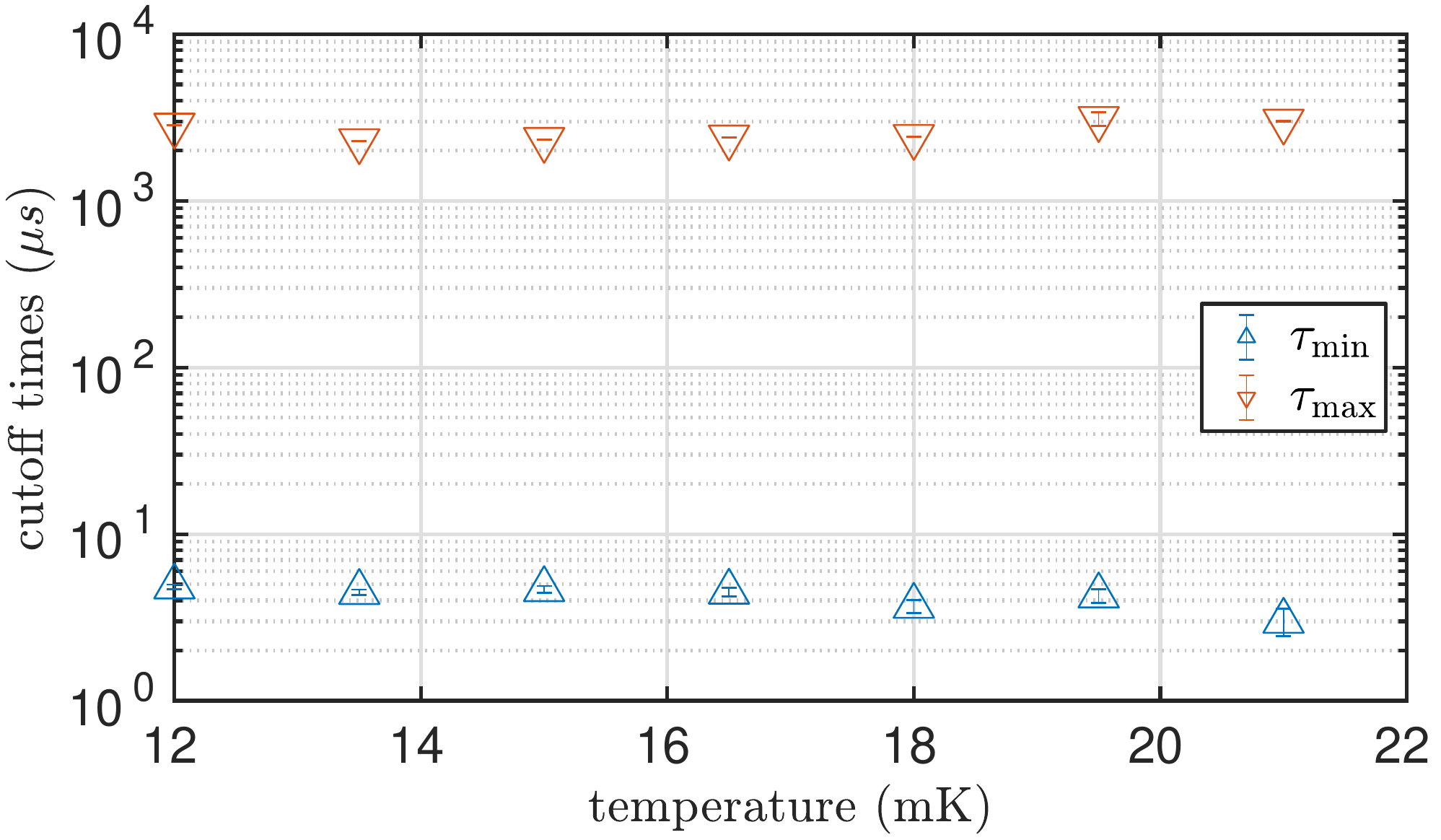}
\includegraphics[width=0.90\linewidth]{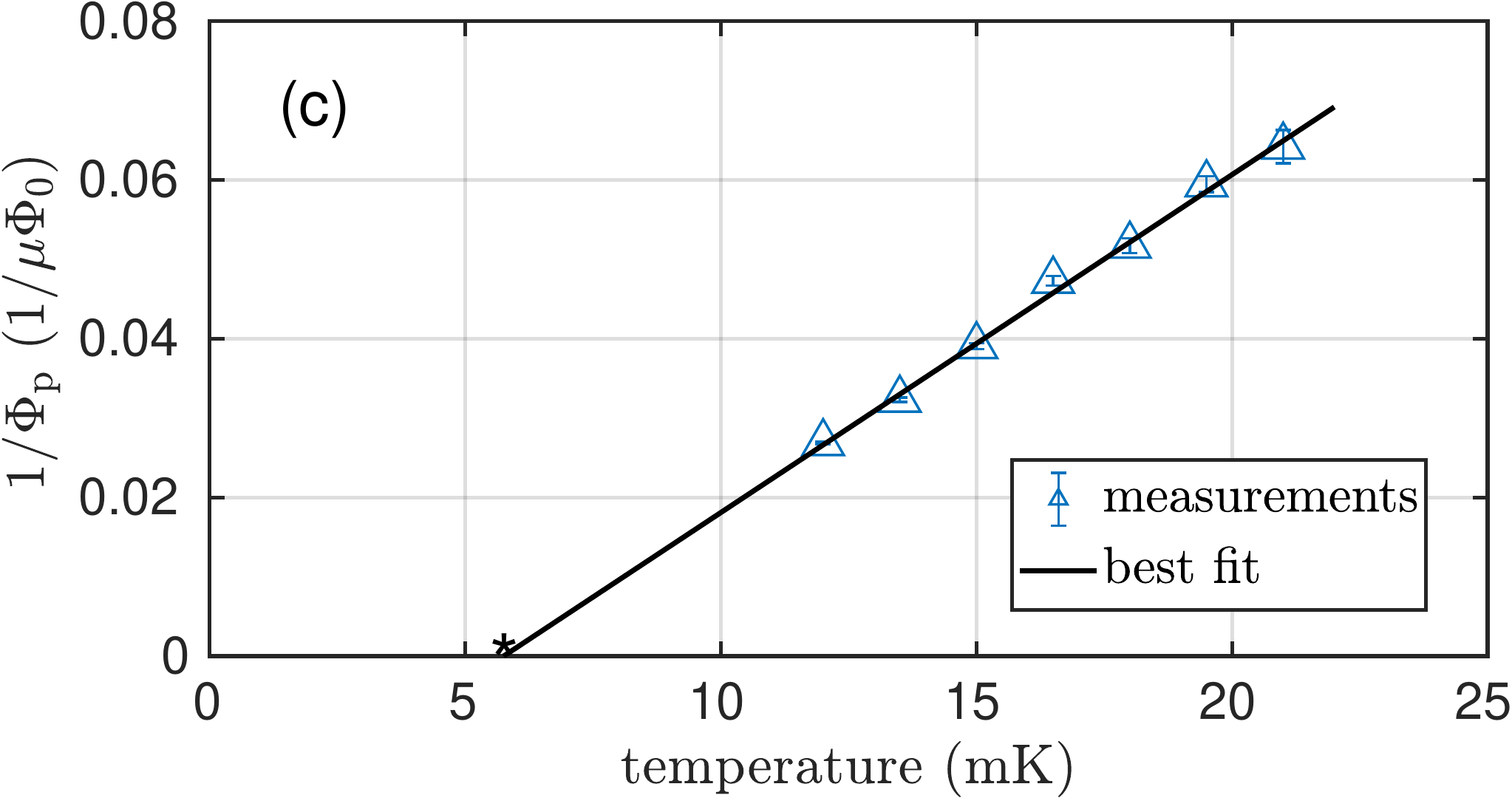}
\caption{(a) Measurements of $\Phi_q^{\rm fb}$ for a range of temperatures. The data shown are from a protocol that varied $\tau_d$ and fixed $\tau_p = 2$ ms. The solid lines show best fits to Eq. \eqref{Eq:DeltaPhi1} using $1/f$ model with parameters $\tau_{\rm min}(T)$ and $\tau_{\rm max}(T)$ shown in (b) and  $\Phi_p(T)$ shown in (c). From the fit to the Curie-Weiss model, we estimate a critical temperature $T_c = 5.7\pm 0.3$ mK.
}
\label{FIG:DepolarizationTemperatureSweep}
\end{figure}

At long polarization times, the theoretical curve, which follows the stretched-exponential law
of Eq.~\eqref{f-sdmodel} with $\nu=1/3$, saturates faster than the experimental data. 
Trying to fit to the homogeneous spin diffusion model results in an even larger deviation due to the exponential law in Eq.~\eqref{f-sdmodel} with $\nu=1$. This suggests that at long times (low frequencies) something beyond spin diffusion is contributing to the flux noise.

Next, we try to fit the data to the empirical $1/f$ based model of Eq.~\eqref{xi-1/f}. Best fits are plotted as solid curves in Fig.~\ref{FIG:SpinPolarizationTime}(a) and (b). The fitting parameters are: 
$\Phi_p = 34.8 \pm 0.1 \, \mu\Phi_0$, $\tau_{\rm  min} = 11.6 \pm 0.1 \, \mu$s, and $\tau_{\rm max} = 8560 \pm 100$ $\mu$s for the polarization curve and $\Phi_p = 35.4 \pm 0.2 \, \mu\Phi_0$, $\tau_{\rm  min} =6.3 \pm 0.2\, \mu$s, and $\tau_{\rm max} = 3470 \pm 5$ $\mu$s for all depolarization curves (averaged over all curves). For all data the best fit $\alpha = 0.98 \pm 0.03$. This model provides a good fit to the experimental data at all timescales, which is not a surprise since it has three fitting parameters; unlike that spin diffusion model that has only one. Nevertheless, the extracted $\alpha$ is close to 1, as expected for $1/f$ noise, and the other fitting parameters are roughly consistent between the polarization curve and all depolarization ones.


We also measured the temperature dependence of the spin environment dynamics by repeating the depolarization experiment at a range of temperatures. Fig.~\ref{FIG:DepolarizationTemperatureSweep}(a) shows typical depolarization data for temperatures ranging from 12.5 mK to 21 mK. We also show best fits of the data to Eq.~\eqref{Eq:DeltaPhi1} with $F(t)$ defined by Eq.~\eqref{xi-1/f}. The cutoff parameters $\tau_{\rm min}$ and $\tau_{\rm max}$ are relatively independent of temperature (Fig.~\ref{FIG:DepolarizationTemperatureSweep}(b)), whereas there is a strong temperature dependence on the polarization amplitude $\Phi_p(T)$ (Fig.~\ref{FIG:DepolarizationTemperatureSweep}(c)). In 
Fig.~\ref{FIG:DepolarizationTemperatureSweep}(c) we also show a fit of the amplitude versus temperature data to the Curie-Weiss model. The best fit to the Curie-Weiss model estimates critical temperature $T_c = 5.7\pm0.3$ mK. The observed proximity to a phase transition is consistent with the previous observation of a $T$-dependent diffusion coefficient in Ref.~\cite{Lanting:2014cm}.

\section{Discussion} 

The experimental results presented here can guide us to a most likely model for flux noise, or at least narrow down the possibilities. Two clear observations stand out that demand explanation:  the initial $\sqrt{t}$ growth of the environmental polarization, and the existence of a phase transition at $T_c \approx 5.7$\,mK. The former is model independent and the latter, although obtained after a fitting, is insensitive to the model; consistent values of $\Phi_p$ were obtained by fitting to different models in Fig.~\ref{FIG:SpinPolarizationTime}.
The close proximity to a phase transition shown in Fig.~\ref{FIG:DepolarizationTemperatureSweep}~(c) 
supports theories that allow for long-range ferromagnetic spin ordering. The $\sqrt{t}$ dependence, on the other hand, is a clear indication of a random diffusion process. The spin diffusion model is consistent with both observations and, as we shall show below, provides a quantitatively consistent description of the observed short-time dynamical behavior above the transition point. Nevertheless, it fails to explain the long-time behavior.

The factor $\Phi_p=\epsilon _{p}/2|I_{p}|$ 
in the polarization/relaxation curves measured in the units of the flux quantum 
$\Phi _{0}$, is identified with the reorganization energy $\epsilon_p$ (the spin polaron
shift, see Eq.~\eqref{Eq:ep}) and therefore is proportional to the static Curie-Weiss
magnetic susceptibility $\chi \left( T\right) =n_{s}\mu _{B}^{2}S\left(
S+1\right) /3\left( T-T_c\right) $, where $\mu _{B}$ is the Bohr magneton, 
$n_{s}$ is the 2D concentration of the interface defects with spin $S$, and $%
T_c$ is the temperature of ferromagnetic phase transition. For a wire interface of width $W$, thickness $h \ll W$, and length of the
loop, $L \gg W$, the factor $\Phi_p$ for spins $S=1/2$ can be expressed as:%
\begin{equation}
\Phi_p\left( T\right) \simeq \frac{2\left( \mu _{0}\mu _{B}\right) ^{2}n_{s}\left\vert I_{p}\right\vert L}{W\left( T-T_c\right) },  \label{A(T)}
\end{equation}%
where $\mu _{0}$ is the magnetic susceptibility of vacuum. From Fig.~\ref{FIG:DepolarizationTemperatureSweep}~(c) it is seen that the
factor $\Phi_p(T)$ clearly obeys the Curie-Weiss law, so that the system of
paramagnetic spins undergoes a ferromagnetic transition at $%
T_c=5.7$~mK. Also, with $\left\vert I_{p}\right\vert =2\mu $A, $L=0.7$%
~mm, and $W=1\mu $m and with the help of Eq. (\ref{A(T)}) we can estimate
the surface spin density as $n_{s}=1.2\times 10^{12}$~cm$^{-2}$ in a
reasonable agreement with a previously reported value of $10^{13} $~cm$^{-2}$ obtained for similar devices~\cite{Lanting:2014cm}. 

To describe the inhomogeneous spin diffusion, we assume that 
$N_{s}$ spins are randomly distributed over a regular
lattice of $N$ sites with the filling factor $x_f=N_s/N$. 
This lattice contains $N_{v}=N-N_{s}=N(1-x_f)$ ``vacancies" (i.e., sites where the spin
is absent) that terminate spin diffusion. The latter is
considered as a process when nonequilibrium magnetization can relax only
via angular momentum transfer between spatially close spins due to 
exchange or dipole-dipole interaction between them. As a result, if the
system is below the percolation threshold with respect to the spin sites,
the diffusion will be confined within finite clusters comprised of connected
nearest-neighbor spin sites that are surrounded by vacancies.

The spin diffusion coefficient ${\cal D}$ 
and the surface spin concentration $n_{s}$
are related because
\begin{equation}
{\cal D}=\eta Ja^{2}, \label{Eq:DaRelation}
\end{equation}
where for a 2D spin environment $a\simeq \left(x_f/n_{s}\right)^{1/2}$ is the distance between the
nearest spins along the direction of the magnetic field (see Fig.~\ref{FIG:SpinPolarizationQubit}(b)), $\eta =\pi ^{1/2}\left( T-T_c\right) /2T$ 
\cite{Kopietz:1998ea}. 
Here $J$ is the effective strength of the
spin-spin coupling, which for the purposes of estimation can be evaluated
as interaction energy of two magnetic spin dipoles with $S=1/2$~\cite{Jackson:1999}: 
\begin{equation}
J\simeq \frac{\mu _{0}\mu _{B}^{2}}{4\pi a^{3}}.  \label{Eq:J}
\end{equation}
In addition, according to the inhomogeneous spin diffusion model (see Appendix \ref{app:spindiffusion} for details), 
the diffusion coefficient ${\cal D}$
is related to the parameter $\Omega$ in Eq.~\eqref{f-sdmodel}  and the average length of the spin cluster 
$\bar w=n_s^{-1/2}(1-x_f^{1/2})^{-1}$ as 
\begin{equation}
\label{Eq:Omega}
\Omega = \frac{\pi^2 {\cal D}}{\bar w^2}. 
\end{equation} 
Using Eqs.~\eqref{Eq:DaRelation}-\eqref{Eq:Omega} with the experimentally extracted values $n_s=1.2\times 10^{12}\text{ cm}^{-2}$ and $\Omega\simeq 100$~Hz at $T=12.5$~mK,
we can estimate the average size of the spin clusters to be $\bar{w} \sim 0.5\, \mu\text{m}$ and 
${\cal D}\simeq 3 \times 10^{-8}\text{ cm}^2/\text{s}$, in agreement with \cite{Lanting:2014cm}. Note that due to the uncertainties in all parameters, these are very rough estimations. Nevertheless, they show consistency among different quantities within the spin diffusion model.

As is clear from Fig.~\ref{FIG:SpinPolarizationTime}(a), spin diffusion predicts faster long time saturation of polarization than the observed data, meaning that the low-frequency noise must have a different origin. This is consistent with the previous results \cite{Lanting:2014cm} indicating that spin diffusion does not explain the $1/f^\alpha$ noise dependence over the observed wide frequency range \cite{Lanting:2009cy,Lanting:2014cm}.
Fitting the polarization and depolarization data to a model based on $1/f^\alpha$ noise provides a nice agreement at all timescales. However, empirical low- and high-frequency cutoffs are needed to achieve a good fit. The fact that $\tau_{\rm  min}$ and $\tau_{\rm max}$ fall within the measured range of $\tau_p$ shows that $1/f^\alpha$ spectrum does not hold over the whole range of relevant frequencies. This was indeed expected, especially at large frequencies, since the short-time $\sqrt{t}$ behavior requires asymptotic $f^{-3/2}$ dependence.

To provide a plausible explanation for these observations, we recall that spin diffusion by construction assumes a constant total magnetic moment. This assumption, although valid at short times, is not expected to hold at long times, especially in the presence of dissipation. Slow evolution of the total magnetic moment can produce additional polarization at long times and contribute to the $1/f^\alpha$ noise spectrum at low frequencies. In an inhomogeneous spin environment, the net magnetic moment of each cluster can slowly grow with time or the clusters can slowly align with the external field, in addition to the changes of their internal magnetic distribution governed by spin diffusion. This demands for a theoretical model that describes both fast spin diffusion dynamics and slow fluctuations of total magnetic moments under a unified framework. It should be mentioned that based on our observations, the environment is above, but close to, the critical temperature, and therefore is in paramagnetic phase. This is in contrast to what some spin cluster models of $1/f^\alpha$ noise assume \cite{kechedzhi2011fractal,de2014ising,de20191}, hence those theories cannot directly apply here.


\section{Conclusion}

We have measured the polarization and relaxation dynamics of magnetic environment coupled to superconducting flux qubits. The extracted equilibrium polarization follows Curie-Weiss temperature dependence, suggesting a ferromagnetic phase transition in the system of environmental spins at a critical temperature $T_c=5.7$~mK. To our knowledge this is the first direct observation of phase transition in the magnetic environment of superconducting devices, although indirect evidences existed before \cite{Sendelbach:2008kr}. 
The measured time dependencies in both the polarization and depolarization experiments are in good agreement with an empirical model that also predicts a noise power spectral density that goes as $1/f^{\alpha}$ for $\alpha \lesssim 1$. We observe a short and long time cutoffs in the spin bath response at $\sim 10$ $\mu$s and $\sim 4$ ms, which correspond to cutoff frequencies $\sim 40$ Hz and $\sim 20$ KHz. This suggests deviation from the  $1/f^{\alpha}$ dependence close to those frequencies. 
An inhomogeneous spin diffusion model with short-time $t^{1/2}$ growth and subsequent stretched-exponential behavior at larger times fits the polarization data up to 1\,ms, but deviates after.
The observed results agree with spin cluster model of the environment. In this picture, fast spin diffusion dynamics within the clusters are responsible for the short-time (high-frequency) response while the slow fluctuations 
produce the $1/f^\alpha$ spectrum 
in the  low frequency regime. The latter may be related to the slow evolution of the magnetic moments of the clusters as a whole. 
More theoretical and experimental investigations are needed to arrive at a more comprehensive model for magnetic flux noise.

\section*{Acknowledgements}

We thank R.~de Sousa and A.~Smirnov for fruitful discussion.

\appendix
\section{Environmental spin polarization}
\label{AppA}

We consider a flux qubit coupled to a spin environment with Hamiltonian
\eqref{Eq:Ham-qubit}.
It can be shown that the relaxation behavior of $\bar \xi(t)$ is tightly connected to the noise spectral density 
\begin{equation} 
S_\xi(\omega) =\frac{1}{2} \int_{-\infty}^\infty dt e^{i\omega t} \langle 
\xi(t)
\xi(0)+\xi(0)\xi(t)
\rangle
\label{Sgamma}
\end{equation}%
through a from of fluctuation dissipation theorem. To see this, let us write the interaction Hamiltonian as
\begin{equation}
\mathcal{H}_{\rm int}(t)= - \xi(t) b(t)
\label{Hint}
\end{equation}%
where $b$ is the force applied to the environmental spins by the qubit, which is proportional to qubit's persistent current. Clearly, $b=\zeta/2$ when the qubit is in classical states with $\sigma_z=\zeta = \pm 1$, and $b=0$ when the qubit is monostable. Using Kubo formula in linear response theory, we have (herein we assume $\hbar=k_B=1$)
\begin{equation} 
\bar \xi(t) = \langle \xi(t) \rangle = i \int_{-\infty}^t dt' \langle [\mathcal{H}_{\rm int}(t'),\xi(t)] \rangle.
\end{equation}%
We have assumed that the expectation $\langle \xi(t) \rangle_0$ at $b=0$ is zero. Introducing retarded Green's function
\begin{equation} 
D(t,t') = D(t-t') = i \langle [\xi(t),\xi(t')] \rangle \theta(t-t'),
\end{equation}%
we obtain
\begin{equation} 
\bar \xi(t) = \int_{-\infty}^\infty dt' D(t-t') b(t'),
\end{equation}%
where we have used the fact that the noise correlations only depend on $t-t'$. Fourier transformation of this equation yields
\begin{equation} 
\bar \xi(\omega) = D(\omega) b(\omega),
\end{equation}%
where 
\begin{equation} 
D (\omega) = \int_0^{\infty} dt e^{i\omega t} D(t),
\end{equation}%
is the frequency dependent susceptibility. Notice that we have used $D(t<0)=0$. From the fluctuation dissipation theorem, we have
\begin{align} 
S_\xi(\omega) &= \coth \left({\omega \over 2T}\right) {\rm Im} D(\omega) \\
&= \coth \left({\omega \over 2T}\right) \int_{0}^\infty dt \sin(\omega t) D(t).
\end{align}%

We consider two cases relevant to our experiments: {\em polarization} and {\em depolarization}. In polarization, $b=0$ from $t=-\infty$ to 0 and is switched on at $t=0$ to $b=\zeta/2$. For $t > 0$, we have
\begin{align}
\bar \xi(t) 
&= -{i\zeta \over 2} \int_0^t dt' \langle [\xi(t'),\xi(t)] \rangle\\
&={i\zeta \over 2} \int_t^0 d(t-t') \langle [\xi(0),\xi(t-t')] \rangle \\
&= {i\zeta \over 2} \int_0^t dt' \langle [\xi(t'),\xi(0)] \rangle,
\end{align}
Therefore
\begin{align}
{d\bar \xi(t)\over dt} ={i\zeta \over 2}\langle [\xi(t),\xi(0)] \rangle = {\zeta \over 2}D(t).
\end{align}
%
%
In depolarization, $b={\zeta/2}$ from $t=-\infty$ and is switched off at $t=0$, we obtain
\begin{align}
{d\bar \xi(t)\over dt} = - {\zeta \over 2}D(t).
\end{align}
Thus, the relaxation function is closely related to the inverse Fourier transform of the susceptibility.

The fluctuation dissipation theorem, for both polarization and depolarization cases, can now be written as
\begin{align}
S_\xi(\omega) &= 2\coth \left({\omega \over 2T}\right) \int_{0}^{\infty } dt\sin(\omega t) \left\vert{d\bar \xi (t) \over dt} \right\vert.
\end{align}%
Expressing in terms of flux noise
\begin{align}
S_\Phi(\omega) &= {1\over 4I_p^2} S_\xi(\omega)\nonumber \\
&= {1\over 2I_p^2}\coth \left({\omega \over 2T}\right) \int_{0}^{\infty }dt\sin(\omega t) \left\vert{d\bar \xi (t) \over dt} \right\vert.
\label{Eq:S_Phi}
\end{align}%
In the classical limit $\omega \ll T$, we have
\begin{align} \label{S(w)}
S_\Phi(\omega) =  {T \over \omega I_p^2} \int_{0}^{\infty }dt\sin(\omega t) \left\vert{d\bar \xi (t) \over dt} \right\vert.
\end{align}%
Note that to obtain $1/\omega$ spectral density, one needs $\bar \xi (t) \sim \log t$. 

\section{Distribution function of the transient order parameter}
\label{Sec:Pgamma} 

If  $\bm B(\bm r)$ represents the magnetic field generated by the qubit at position $\bm r$ and $\bm M(\bm r,t)$ is the magnetization of the environment at the same point and at time $t$, then 
\begin{equation} 
\xi(t) =2 \zeta\int d\bm r\bm B(\bm r)\cdot \bm M(\bm r,t),  \label{gamma}
\end{equation}%
For simplicity, throughout the rest of the paper we only consider $\zeta = +1$. The order parameter $\xi$ is proportional to the flux  through the qubit that is generated by the environment: $\xi =2I_{p} \,\delta \Phi _{q}^{\mathrm{se}}$, where $I_p$ is the qubit's persistent current. In equilibrium, we have  $\bm M(\bm r,t\to \infty) = \chi \bm B(\bm r)$, where $\chi$ is the magnetic susceptibility, therefore 
\begin{equation}
\bar \xi(t \to \infty) = 2 \chi \int d\bm r\bm B^2(\bm r) = \epsilon_p,
\end{equation} 
where $\epsilon_p$ is the equilibrium reorganization energy.
To study dynamics of the spin environment, we use Landau-Ginsburg Hamiltonian: 
\begin{eqnarray}
\mathcal{H}_{en} = \int\!\! d{\bm r}\left[a {\bm M}({\bm r},t)^2 - {\bm B}({\bm r})\cdot {\bm M}({\bm r},t) \right. \nonumber \\
 \left. + b (\nabla{\bm M})^2 + c {\bm M}^4 + \dots \right].
\end{eqnarray}
The coefficient $a$ is related to the static magnetic susceptibility $\chi$ as $a=1/(2\chi)$. Since
our primary goal is to describe dynamical effects related to spin diffusion in a paramagnetic phase we may omit the gradient term and various fourth-order terms that are not crucially important for our purpose, and concentrate on a simplified second-order Hamiltonian: 
\begin{equation}  \label{Eq:Hamiltonian}
\mathcal{H}_{en}= \sum_{\alpha=x,y,z} \int d{\bm r} \left[\frac{M_\alpha^2(\bm r,t)}{2\chi}-B_\alpha(%
\bm r) M_\alpha(\bm r)\right]
\end{equation}
Here $M_\alpha$ and $B_\alpha$ are magnetization and
external magnetic field, respectively. We also assume that magnetization is a conserved quantity satisfying the continuity equation: 
\begin{equation}  \label{Eq:continuity}
\frac{\partial M_\alpha(\bm r,t)}{\partial t}+\nabla \cdot {\bm j}_\alpha=0.
\end{equation}
Here ${\bm j}_\alpha$ is the magnetization (spin) current, which can be
calculated as: 
\begin{equation}  \label{Eq:current}
{\bm j}_\alpha(\bm r,t)=-\xi{\bm \nabla}\frac{\delta \mathcal{H}_{en}}{\delta
M_\alpha},
\end{equation}
where $\xi$ is the Onsager transport coefficient~\cite%
{Bennett:1965uj,Chaikin:2009ej}. Using Eqs.~\eqref{Eq:Hamiltonian} and Eq.~\eqref{Eq:current} along with the
Einstein relation $\xi =\mathcal{D}\chi $ and substituting Eq.~%
\eqref{Eq:current} into \eqref{Eq:continuity} yields the following diffusion
equation for the magnetization component $M_{\alpha }({\bm r},t)$: 
\begin{equation}
\frac{\partial M_{\alpha }(\bm r,t)}{\partial t}=\mathcal{D}\nabla ^{2}\left[
M_{\alpha }(\bm r,t)-{\chi }B_{\alpha }(\bm r)\right] .  \label{Eq:diffusion}
\end{equation}%

Suppose we know the eigenfunctions of the stationary diffusion equation
(which form coincides with that of the Schr{\"{o}}dinger equation) 
\begin{equation}
\mathcal{D}\nabla ^{2}\varphi _{n}(\bm r)=-\frac{\varphi _{n}(\bm r)}{\tau
_{n}},  \label{Eq:SE}
\end{equation}%
where $n$ enumerates diffusion modes. We expand quantities $M_{\alpha }(\bm r,t)$ and $B_{\alpha }(\bm r)$ using the complete set of orthonormal functions $\{\varphi _{n}(\bm r)\}$: 
\begin{align}
M_{\alpha }(\bm r,t)& =\sum_{n}\mu _{\alpha n}(t)\varphi _{n}(\bm r),
\label{Eq:Mt} \\
B_{\alpha }(\bm r)& =\sum_{n}B_{\alpha n}\varphi _{n}(\bm r). \label{Eq:Bn}
\end{align}%
Now we can substitute Eqs.~\eqref{Eq:Mt} and \eqref{Eq:Bn} in the diffusion
equation~\eqref{Eq:diffusion} and obtain a set of kinetic (Langevin)
equations for each diffusion mode $\mu _{\alpha n}$: 
\begin{equation}
\dot{\mu}_{\alpha n}=-\frac{\mu _{\alpha n}-\chi B_{\alpha n}}{\tau _{n}}%
+\delta f_{\alpha n}(t),  \label{Eq:Langevin}
\end{equation}%
where $\delta f_{\alpha n}$ is a $\delta $-correlated Gaussian white noise
(random force), chosen to ensure the fulfillment of the
fluctuation-dissipation theorem. Accordingly, the Hamiltonian in Eq.~\eqref{Eq:Hamiltonian} can be represented as a sum of the
individual-mode Hamiltonians $\mathcal{H}_{en}=\sum_{\alpha n}\mathcal{H}_{\alpha
n}-\epsilon _{p}/2$, where 
\begin{equation}
\mathcal{H}_{\alpha n}=\frac{1}{2\chi }\left( \mu _{\alpha n}-\chi B_{\alpha
n}\right) ^{2}  \label{Eq:Hn}
\end{equation}%
and 
\begin{equation}
\epsilon _{p}=2\chi \sum_{\alpha n}B_{\alpha n}^{2}.  \label{Eq:ep_append}
\end{equation}%
Eqs.~\eqref{Eq:Langevin} and~\eqref{Eq:Hn} resemble Brownian motion of a particle with mass $M=1/\chi$ moving with velocity $v =\mu _{\alpha n}-\chi B_{\alpha n}$ and having damping (relaxation) rate $\gamma = 1/\tau_n$.
We can therefore construct a Fokker-Plank equation for the probability density $P(v,t)$: 
\begin{equation}
\frac{\partial P}{\partial t}=\gamma \frac{\partial }{%
\partial v}\left( v P + {T \over M} \frac{\partial P}{\partial v}\right).  \label{Eq:Fokker}
\end{equation}%
It can be checked (by substitution) that the solution to this equation is
\begin{equation}
P(v,t) =\frac{1}{(2\pi T/M)^{1/2}}\exp \left[ -\frac{(v-v_0e^{-\gamma t} )^{2} }{2 T/M}\right],
\label{Eq:P(v,t)}
\end{equation}%
where $v_0$ is the expectation of $v$ at $t=0$. 

For the polarization situation where the qubit's persistent current is zero ($B_{\alpha n}=0$) for $t<0$ and is turned on at $t=0$, we have $\langle \mu_{\alpha n} \rangle =0$, thus, $v_0 = -\chi B_{\alpha n}$. Therefore, for each diffusion mode we have
\begin{equation}
P_{n}^{\alpha }\!\left( \mu _{\alpha n},t\right) =\frac{1}{(2\pi \chi
T)^{1/2}}\exp \left[ -\frac{(\mu _{\alpha n}{-}(1{-}e^{-t/\tau_n})B_{\alpha n}\chi )^{2}}{2\chi T}\right].  \label{Eq:Pn2}
\end{equation}%
The order parameter (\ref{gamma}) in this representation becomes 
\begin{equation}
\xi =2\sum_{\alpha n}B_{\alpha n}\mu _{\alpha n}  \label{Eq:gamma_def}
\end{equation}%
This allows us to find the probability distribution of $\xi $ as a
function of polarization time $t$: 
\begin{equation}
\mathcal{P}(\xi ,t)\!\!=\!\!\!\!\int \!\!\prod_{\alpha n}d\mu _{\alpha
n}\delta \left( \xi -2\sum_{\alpha n}B_{\alpha n}\mu _{\alpha n}\right)
P_{n}^{\alpha }\left( \mu _{\alpha n},t\right)  \label{Eq:Pgamma-product}
\end{equation}%
By using Eq.~\eqref{Eq:Pn2} we can calculate the multiple Gaussian integral
in Eq.~\eqref{Eq:Pgamma-product} in a standard way as follows. First, we employ the
Fourier transform to remove the constraint imposed by the $\delta $%
-function: 
\begin{equation*}
\delta ( \xi -2\sum_{n}B_{\alpha n}\mu _{\alpha n} ) =\frac{1}{%
2\pi }\int_{-\infty }^{\infty }dke^{ik\left( \xi -2\sum_{\alpha
n}B_{\alpha n}\mu _{\alpha n}\right) }
\end{equation*}%
and then perform series of simple Gaussian integrations to obtain: 
\bigskip 
\begin{align}
\mathcal{P}(\xi ,t)
 =\frac{1}{\sqrt{4\epsilon _{p}T}}\exp \left[ -\frac{(\xi -\bar\xi
(t))^{2}}{4\epsilon _{p}T}\right], \label{Eq:Pgamma}
\end{align}
where
\begin{align}
\bar\xi (t) &= 2\chi \sum_{\alpha n}B_{\alpha n}^{2}\left( 1{-}e^{-t/\tau
_{n}}\right) \nonumber \\
&= \epsilon_{p} \left[ 1 - \sum_{n}p_{n}e^{-t/\tau _{n}} \right],
\label{Eq:gamma(t)}
\end{align}%
is the ensemble average of $\xi$ at time $t$, $\epsilon_p$ is the reorganization energy given by Eq.~\eqref{Eq:ep_append}, and  
$p_n = \sum_{\alpha}B_{\alpha n}^{2}/\sum_{\alpha,n}B_{\alpha n}^{2}$ measures the relative contribution of the $n$th diffusion mode in the relaxation process. 

For the case of depolarization, the qubit's persistent current is nonzero for $t<0$ and is turned off at $t=0$ by making the qubit monostable. The environment is therefore polarized to $\mu^0_{\alpha n}$ leading to the initial value of the order parameter 
\begin{equation}
\bar \xi (0) =2\sum_{\alpha n}B_{\alpha n}\mu^0_{\alpha n}.
\end{equation}%
If the polarization time is $t_0$, then $\mu^0_{\alpha n}=\chi B_{\alpha n}(1-e^{-t_0/\tau_n})$. At $t>0$, qubit persistent current is absent, hence $B_{\alpha n}=0$. The initial velocity in Eq.~\eqref{Eq:P(v,t)} is therefore $v_0 =\mu^0_{\alpha n}$, leading to
\begin{equation}
P_{n}^{\alpha }\!\left( \mu _{\alpha n},t\right) =\frac{1}{(2\pi \chi
T)^{1/2}}\exp \left[ -\frac{(\mu _{\alpha n}{-}\mu^0_{\alpha n}e^{-t/\tau_n})^{2}}{2\chi T}\right].  \label{Eq:Pndepol}
\end{equation}%
The order parameter $\bar \xi(t)$ is defined in the bistable state of the qubit according to \eqref{Eq:Ham-qubit}. We therefore use \eqref{Eq:gamma_def} for its definition, keeping in mind that $B_{\alpha n}$ correspond to the bistable state of the qubit. In other word, $\bar \xi (t)$ is the energy bias the qubit would experience if it becomes bistable at time $t$.
The probability distribution of $\xi $ is again given by \eqref{Eq:Pgamma-product}
with $P_{n}^{\alpha }\!\left( \mu _{\alpha n},t\right)$ defined in Eq.~\eqref{Eq:Pndepol}. Following the same calculations as before, we arrive at \eqref{Eq:Pgamma} with 
\begin{align}
\bar\xi (t) &= 2 \sum_{\alpha n}B_{\alpha n}\mu^0_{\alpha n}e^{-t/\tau_n} \nonumber \\
& = 2 \chi \sum_{\alpha n}B_{\alpha n}^2 (1-e^{-t_0/\tau_n})e^{-t/\tau_n}  \nonumber \\
& = \epsilon_p \sum_{\alpha n}p_n [e^{-t/\tau_n}-e^{-(t+t_0)/\tau_n}]
\label{Eq:gamma2(t)}
\end{align}%

Equations \eqref{Eq:gamma(t)} and  \eqref{Eq:gamma2(t)} can be written as
\begin{align}
\bar\xi (t) = \epsilon_{p} [ 1 - F(t) ],
\label{polxi}
\end{align}%
for polarization, and 
\begin{equation}
\bar\xi (t) = \epsilon_{p} [F(t)-F(t+t_0)].
\label{depolxi0}
\end{equation}%
for depolarization, where $t_0$ is the polarization time during the polarization process and
\begin{equation}
F(t)= \sum_{n}p_{n}e^{-t/\tau _{n}}.
\label{F(t)}
\end{equation}%
 Notice that $F(0)=1$ and $F(\infty) = 0$. Therefore, the initial polarization in \eqref{depolxi0} is 
$\bar\xi (0) = \epsilon_{p} [1-F(t_0)]$, in agreement with $t_0$ being the polarization time. Equations \eqref{Eq:Pgamma}-\eqref{depolxi0}, although derived for the fluctuations of spin diffusion modes, hold for any set of independent fluctuators following Langevin dynamics. 

Substituting Eq.~\eqref{Eq:gamma(t)} or \eqref{Eq:gamma2(t)} into Eq.~\eqref{S(w)}, we obtain
\begin{align} \label{Sw2}
S_\Phi(\omega) &=  {\epsilon _{p}T \over \omega I_p^2} \int_{0}^{\infty }dt\sin(\omega t)\sum_{n}{p_{n} \over \tau_n} e^{-t/\tau _{n}} \nonumber \\
&=  {\epsilon _{p}T \over \omega I_p^2}\sum_{n}{p_{n} \over \tau_n} \int_{0}^{\infty }dt\sin(\omega t) e^{-t/\tau _{n}} \nonumber \\
&=  {\epsilon _{p}T \over I_p^2}\sum_{n}  {p_{n}\tau_n \over \omega^2 \tau_n^2 +1}.
\end{align}%
One may also take the continuous-$\tau$ limit of \eqref{Sw2} by replacing $p_n$ with the distribution $p(\tau)$:
\begin{align}
S_\Phi(\omega) &= {\epsilon _{p}T \over I_p^2}\int_0^\infty d\tau {p(\tau)\tau \over \omega^2 \tau^2 +1}.
\end{align}%

We now derive $F(t)$ for a few different models. For simplicity, we only consider polarization cases in detail.

\section{$1/f^\alpha$-noise model} 
\label{app:logarithmic}
To achieve $1/f^\alpha$ noise spectrum, we need
\begin{equation}
p(\tau ) = \left\{ 
\begin{array}{cc}
{\cal N}\tau^{\alpha-2} & \ \ \tau_{\text{min}} < \tau < \tau_{\text{max}} \ \\
0 & \ \ \text{otherwise}
\end{array}%
\right.  \label{p(tau)2}
\end{equation}%
where 
\begin{equation}
{\cal N}^{-1} = \left\{ 
\begin{array}{cc}
 \log( \tau _{\text{max}}/\tau _{\text{min}})  & \qquad \alpha = 1 \ \\
(\alpha{-}1)^{-1} [\tau_{\text{max}}^{\alpha-1} - \tau_{\text{min}}^{\alpha-1}] & \qquad  \alpha \ne 1
\end{array}%
\right. 
\end{equation}%
is a normalization factor. Substituting \eqref{p(tau)2}  into \eqref{Sw2}, we find $S_\Phi(\omega) \sim \omega^{-\alpha}$ for $\tau_{\rm max}^{-1} {<} \omega {<} \tau_{\rm min}^{-1}$. 
Also, from \eqref{F(t)}, we obtain
\begin{align}
F(t) &=\int_0^\infty d\tau p(\tau) e^{-t/\tau} \nonumber \\
&= {\cal N} \int_{\tau_{\text{min}}}^{\tau_{\text{max}}} d\tau \, \tau^{\alpha-2} e^{-t/\tau} \nonumber \\
&= {\cal N} t^{\alpha-1} \int_{t/\tau_{\text{max}}}^{t/\tau_{\text{min}}} du \, u^{-\alpha} e^{-u} \nonumber \\
& = {\cal N} t^{\alpha-1} [\Gamma(1{-}\alpha , \, t/\tau_{\text{max}}) - \Gamma(1{-}\alpha , \, t/\tau_{\text{min}})].
\label{Eq:Phi2}
\end{align}%
where 
\begin{equation}
\Gamma \left(s,x\right) = \int_x^\infty t^{s-1}e^{-t}dt
\end{equation}
is the incomplete gamma-function.

\section{Spin diffusion model}
\label{app:spindiffusion}
We now move to more elaborate theories based on spin diffusion. Consider the simplest case of environmental spins in a thin wire of length $L$ and width $W$. We neglect the height of the wire and assume that the flux noise is produced by environmental spins on the 2D interface of the wire. If $x$ and $z$ represent directions along the width and length of the wire respectively, the magnetic field generated by the persistent current,  $B(x)$, will only be a function of $x$ and independent of $z$. The behavior of the homogeneous system is therefore effectively 1D. We can therefore calculate the contribution of a narrow region with width $dz$ to the order parameter and then add them up. Such a narrow region would essentially behave like a spin chain. We consider the homogeneous spin diffusion model as a special case of the inhomogeneous model, where only one length scale is associated with all chains. For the inhomogeneous case, we assume there are vacancies (defects) that break the chain into smaller 1D regions of length $w_i$, with $\sum_i w_i = W$. 

For the $i$th region, the solutions, $\varphi_n(x)$, to Eq.~\eqref{Eq:SE} are Fourier terms $\sin k_n x$ and  $\cos k_n x$, with $k_n=2\pi n/w_i$ and $\tau_n^{-1} = \gamma_n = \mathcal{D} k_n^2$. We can approximate the magnetic field  in the $i$th segment
\begin{equation}
B(x) = \bar B_i+B'_i \, x, \quad
-w_i/2 \leq x \leq w_i/2,
\end{equation}
where $\bar B_i$ is the average magnetic filed and $B'_i = (\partial B/\partial x)_0$. Fourier expansion of $B$ is given by
\begin{equation}
B(x) - \bar B_i=\sqrt{\frac{2}{w_i}}\sum\limits_{n=1}^{\infty }B^i_{n}\sin k_n x
\label{Eq:B-Fourier}
\end{equation}%
where the cosine terms are absent by construction and the form-factors $B^i_n$
can be expressed as:
\begin{eqnarray}
\label{Eq:Bn-new}
B^i_{n} &=& \sqrt{\frac{2}{w_i}}\int_{-w_i/2}^{w_i/2}dx B(x)\sin k_n x \nonumber \\
&=& B'_i\frac{(-1)^{n+1}\sqrt{2w_i}}{k_n}.
\end{eqnarray}

The magnetization is also independent of $z$ and similar to $B(x)$ has the Fourier expansion
\begin{equation}
\label{Eq:Fourier}
M(x,t)=\sqrt{\frac{2}{w_i}}\sum_{n=1}^{\infty} \mu_n(t)\sin k_n x.
\end{equation}
Notably, the term with $n=0$ is absent in Eq.~\eqref{Eq:Fourier}. This is because the total magnetization is a conserved quantity and is assumed to be zero ( $\int dx M(x,t)=0$) before the magnetic field was turned on, in the polarization case, and will remain zero despite the presence of $\bar B$. If the only relaxation mechanism in the system is spin diffusion the total magnetic moment will remain zero although the local magnetization will be induced by the nonuniform components of the magnetic field. Uniform magnetic field may induce magnetization only in the presence of some local relaxation mechanism, e.g., spin-phonon relaxation.

We can now calculate contribution of this region to the order parameter. From \eqref{Eq:gamma(t)}, we have
\begin{align} 
d \bar \xi_i(t) &=2dz\int dx B(x)M(x,t) \nonumber \\
&= 2 dz \chi \sum_{n=1}^{\infty} B_{n}^{i\,2} ( 1-e^{-t/\tau _{n}}),  
\end{align}%
where  $\tau _{n}^{-1}={\cal D}\left( 2\pi n/w_i\right) ^{2}$. Substituting $B^i_n$ by \eqref{Eq:Bn-new}, we obtain
\begin{align}  \label{dxi}
d \bar \xi_i(t) &= dz \chi B_i^{\prime \, 2}
\sum_{n=1}^{\infty}  \frac{w_i^3}{\pi^2n^2}( 1-e^{-{\cal D}\left( 2\pi n/w_i\right) ^{2} t}).  
\end{align}
Assuming that the size of clusters are distributed according to the distribution $P(w)$, the order parameter is given by 
\begin{align}
\bar\xi(t) =& L \chi \sum_i B_i^{\prime \, 2} \sum_{n=0}^{\infty }\int_{0}^{\infty } dw P(w)  \notag  \label{xiPw} \\
& \times  \frac{w^3}{\pi ^{2}n^{2}} \left[ 1-e^{-{\cal D}\left( 2\pi n/w\right) ^{2} t}
\right].
\end{align}%
We now need to find $P(w)$, which we shall do for both inhomogeneous and homogeneous cases.

\subsection{Inhomogeneous case}
\label{app:inhomogeneousspindiffusion}
Let $\eta$ be the linear density of defects. For a region of length $w$, the average number of defect is $\eta w$. The probability of having $k$ defects within this region is given by Poisson distribution: $e^{-\eta w} (\eta w)^k/k!$, hence the probability of this region being defect-free ($k=0$) is $e^{-\eta w}$. The distribution of $w$ is therefore given by the normalized probability density 
\begin{equation}
P(w) = \bar w^{-1} e^{-w/\bar w},
\end{equation}
where $\bar w = \eta^{-1}$ is the average length of the clusters.
The expected contribution of a region can be calculated by integrating \eqref{dxi} over all $w$ with the above distribution. Summing over the whole wire, we obtain
\begin{align}
\bar\xi(t) =& L \chi \sum_i B_i^{\prime \, 2} \sum_{n=0}^{\infty }\int_{0}^{\infty }\frac{dw}{\bar w\pi ^{2}n^{2}} e^{-w/\bar w}  \notag  \label{Eq:dgamma1} \\
& \times w^{3}  \left[ 1-e^{-{\cal D}\left( 2\pi n/w\right) ^{2} t}
\right].
\end{align}%
To evaluate the sum in Eq.~\eqref{Eq:dgamma1} we change the
variables in each term of the series by replacing $w$ with $2n\bar wv$ and
interchanging the order of summation and integration. This yields: 
\begin{align}
\bar \xi (t) =& \frac{16L \bar w^3\chi }{\pi ^{2}}\sum_i B_i^{\prime \, 2}\int_{0}^{\infty }dv v^{3} \left(1{-}e^{-\Omega t/v^{2}}\right) \sum_{n=0}^{\infty
}n^{2}e^{-2nv }  \label{Eq:dgamma2}
\end{align}
where $\Omega = \pi^2{\cal D}/\bar w^2$. Note that $\sqrt{{\cal D}t}$ is the length over which spin diffusion happens after time $t$, therefore $\Omega^{-1}$ is the timescale for diffusion to reach the length $\bar w$.
Calculating the sum in Eq.~\eqref{Eq:dgamma2} is straightforward and we
finally obtain
\begin{equation}\label{xifinal}
\bar\xi(t) =\epsilon_p \left[ 1-F(t)\right], 
\end{equation}
where $\epsilon_p =L \bar w^3\chi  \sum_i B_i^{\prime \, 2}$ and
\begin{equation}
F(t)=\frac{4}{\pi ^{2}}\int_{0}^{\infty }\!\!\!dv v^{3}\frac{\coth (v
	)}{\sinh ^{2}(v)} \, e^{-\Omega t/v^{2}}. \label{Eq:F(t)}
\end{equation}
Note that $F(0)=1$ and $F(\infty)=0$. Therefore, $\bar\xi(\infty) = \epsilon_p$, as expected.

The noise spectral density based on this model is 
\begin{align}
S_\Phi(\omega) &=  {T \over \omega I_p^2} \int_{0}^{\infty }dt\sin(\omega t) \left\vert{d\bar \xi (t) \over dt} \right\vert. \nonumber \\
&=  {4\Omega T \epsilon_p \over \omega \pi^{2}I_p^2} \int_{0}^{\infty }\!\!\!dv v \frac{\coth (v)}{\sinh ^{2}(v)} \int_{0}^{\infty }dt\sin(\omega t) e^{-\Omega t/v^{2}}. \nonumber \\
&=  {4T \epsilon_p \over \Omega \pi^{2}I_p^2} \int_{0}^{\infty }\!\! dv  \frac{\coth (v)}{\sinh ^{2}(v)} {v \over (\omega/\Omega)^2 + v^{-4}}. \nonumber \\
\end{align}%

It is instructive to investigate the short and long-time asymptotics of $\bar\xi (t)$. For $\Omega t \ll 1$, or equivalently $\sqrt{{\cal D}t} \ll \bar w$, we consider
\begin{equation}
1- F(t)=\frac{4}{\pi ^{2}}\int_{0}^{\infty }\!\!\!dv v^{3}\frac{\coth (v
	)}{\sinh ^{2}(v)} (1-e^{-\Omega t/v^{2}}), 
\end{equation}
which is well behaved in both small and large $v$ integration limits. The dominant contribution to the integral comes form small $v$ regions, for which we can approximately write
\begin{equation}
\frac{\coth (v)}{\sinh^{2}(v)} \approx v^{-3}.
\end{equation}
Substituting back and changing the integration variable to $u=K/u$, we obtain
\begin{equation}
1- F(t)=\frac{4\sqrt{\Omega t}}{\pi ^{2}}\int_{0}^{\infty } {du \over u^2} (1-e^{-u^{2}}) = \frac{16\sqrt{\Omega t}}{\pi ^{7/2}}. 
\end{equation}
Therefore, 
\begin{equation}
F(t) \approx 1-C\sqrt{\Omega t}, 
\end{equation}
where $C=16\pi^{-7/2}$. This means for short times $\bar\xi(t) \sim \sqrt{t}$, similar to the homogeneous case discussed in the next subsection. The reason is when the spin diffusion length, $\sqrt{{\cal D}t}$, is much smaller than the average length of the clusters, $\bar w$, disorder is effectively invisible to the diffusion process and the system should behave similar to a homogeneous spin system.

When $\Omega t \gg 1$, the integral in \eqref{Eq:F(t)} is dominated by large $v$ regions. We can therefore substitute
\begin{equation}
\frac{\coth (v)}{\sinh^{2}(v)} \approx 4 e^{-2v}
\end{equation}
into \eqref{Eq:F(t)} to obtain
\begin{equation}
F(t)=\frac{16}{\pi ^{2}}\int_{0}^{\infty } \! dv v^{3}e^{-2v-{\Omega t/{v^{2}}}}.
\end{equation}
The integrand is sharply peaked near $v_0=(\Omega t)^{1/3}$. Using the
steepest descent method, we substitute $v=v_0$ in the exponent to obtain
\begin{equation}
F(t) \sim e^{-3v_0} = e^{-3(\Omega t)^{1/3}}.
\end{equation}
The timescale $\Omega^{-1}$ is the time needed for spin polarization to diffuse over a length of the order of the average size of the clusters, $\bar w$.
The short and long-time asymptotic behavior of $F(t)$ can therefore be summarized by
\begin{equation} 
F(t) = \left\{ \begin{array}{cl} 
 1- C \sqrt{\Omega t} & \text{  for } t \ll \Omega^{-1} \\
C^\prime  e^{-3(\Omega t)^{1/3}} & \text{  for } t \gg \Omega^{-1}
\end{array}
\right. .
\end{equation}%

\subsection{Homogeneous case}

In the homogeneous case, there is only one length scale $\bar w$ for all chains, determined by the geometry. Therefore
\begin{equation}
P(w) = \delta(w-\bar w),
\end{equation}
where $\bar w = O(W)$, with $W$ being the width of the wire.
Substituting into \eqref{xiPw}, we obtain
\begin{align} \label{xiHomo0}
\bar\xi(t) =& L \chi B^{\prime \, 2} \sum_{n=0}^{\infty } \frac{\bar w^3}{\pi ^{2}n^{2}}    \left[ 1-e^{-{\cal D}\left( 2\pi n/\bar w\right) ^{2} t}
\right].
\end{align}%
Using $\bar\xi(t {\to} \infty) = \epsilon_p$, we have
\begin{align} 
\label{Eq:ep}
\epsilon_p =  L \chi B^{\prime \, 2} \frac{\bar w^3}{\pi ^{2}} \sum_{n=1}^{\infty } \frac{1}{n^{2}} = {1 \over 6} L \chi B^{\prime \, 2} \bar w^3.   
\end{align}%
Therefore, \eqref{xiHomo0} can be written as
\begin{align} \label{xiHomo1}
\bar\xi(t) = \epsilon_p [1-F(t)],
\end{align}%
where
\begin{align} \label{FtHomo1}
F(t) = {6 \over \pi^2} \sum_{n=1}^{\infty } \frac{e^{- 4(\Omega t)n^2}}{n^{2}} ,
\end{align}%
with $\Omega = {\cal D}\left(\pi/\bar w\right) ^{2}$, defined the same as in the inhomogeneous case.


We can now find the short- and long-time asymptotic behavior of $F(t)$. For long timescales, only $n=1$ survives in the exponential of \eqref{xiHomo1}. Hence 
\begin{align} 
F(t) = C^\prime e^{- 4\Omega t},
\end{align}%
where $C^\prime = 6\pi^{-2}$.
For short timescales we can replace the sum with an integral
\begin{align}
1-F(t) =& {6 \over \pi^2} \int_{0}^{\infty } \frac{dv}{v^{2}}    \left[ 1-e^{- 4\Omega t  v^2}
\right]
\end{align}%
Changing the integration variable to  $u = 2\sqrt{\Omega t}\, v$, we obtain
\begin{align}
1-F(t) =& {6 \over \pi^2} \sqrt{\Omega t} \int_{0}^{\infty } \frac{du}{u^{2}}    \left[ 1-e^{-u^2} \right] \nonumber \\
=& {48 \over \pi^{7/2}} \sqrt{\Omega t} .
\end{align}%
The short- and long-time asymptotic behavior of $F(t)$ can therefore be summarized by
\begin{equation} 
F(t) = \left\{ \begin{array}{cl} 
1- C \sqrt{\Omega t} & \text{  for } t \ll \Omega^{-1} \\
C^\prime e^{-4\Omega t} & \text{  for } t \gg \Omega^{-1}
\end{array}
\right. ,
\end{equation}%
where $C = 48\pi^{-7/2}$ and $C^\prime=6\pi^{-2}$. The $\sqrt{t}$ dependence at short times leads to $S_\Phi(\omega) \sim \omega^{-3/2}$ at large frequencies, above $f_c \sim W^2/{\cal D}$, in agreement with the numerical calculations in Ref.~\cite{Lanting:2014cm}. The crossover frequency in  Ref.~\cite{Lanting:2014cm} is 0.1-1 Hz, which means the  $\sqrt{t}$ dependence is expected to continue up to 1-10 s.

\bibliographystyle{apsrev}
\bibliography{main}

\end{document}